\documentclass[twocolumn,final,aps,prl,showpacs,superscriptaddress,amsmath,amssymb,amsfonts,floatfix]{revtex4-1}

\usepackage{graphicx} 
\usepackage{booktabs}
\usepackage{multirow}
\usepackage{makecell} 
\usepackage{array}
\usepackage{url}
\usepackage{amsmath}   
\usepackage{siunitx}  
\usepackage{float}
\usepackage[normalem]{ulem}

\usepackage[colorlinks,breaklinks,bookmarks=true,citecolor=blue,linkcolor=blue,urlcolor=blue]{hyperref}
\usepackage{color}
\usepackage{tabularx}
\usepackage{sidecap}
\usepackage{soul}
\usepackage{color}
\usepackage{orcidlink}

\usepackage{soul}
\usepackage{ulem}

\begin{document}

\author{Ao Zhang\,\orcidlink{0000-0001-9786-3942}}
\affiliation{School of Physics, Southeast University, Nanjing 211189, China}
\affiliation{School of Physics, Northwest University, Xi'an 710127, China}

\author{Chao Deng\,\orcidlink{0009-0003-8433-2909}}
\affiliation{School of Physics, Northwest University, Xi'an 710127, China}

\author{Wenfeng Wu\,\orcidlink{0000-0002-6575-5813}}
\affiliation{Institute of Solid State Physics, TU Wien, 1040 Vienna, Austria}

\author{Run Yang\,\orcidlink{0000-0002-9547-756X}}
\email{ryang@seu.edu.cn}
\affiliation{School of Physics, Southeast University, Nanjing 211189, China}
\affiliation{Jiangsu Key Laboratory of New High Temperature Superconducting Materials for Controllable Nuclear Fusion, Southeast University, Nanjing 211189, China}

\author{Zhixiang Shi\,\orcidlink{0000-0003-3881-5152}}
\email{zxshi@seu.edu.cn}
\affiliation{School of Physics, Southeast University, Nanjing 211189, China}
\affiliation{Jiangsu Key Laboratory of New High Temperature Superconducting Materials for Controllable Nuclear Fusion, Southeast University, Nanjing 211189, China}

\author{Liang Si\,\orcidlink{0000-0003-4709-6882}}
\email{siliang@nwu.edu.cn}
\affiliation{School of Physics, Northwest University, Xi'an 710127, China}
\affiliation{Shaanxi Key Laboratory for Theoretical Physics Frontiers, Xi'an 710127, China}
\affiliation{Fundamental Discipline Research Center for Quantum Science and technology of Shaanxi Province, Xi'an 710127, China}

\title{Minimal Three-Band Model and $d$-Wave Superconductivity in La$_4$Ni$_3$O$_8$}

\begin{abstract}
Motivated by the recent observation of ambient-pressure superconductivity in Pr- and Y-substituted La$_4$Ni$_3$O$_8$, we combined density-functional theory (DFT), dynamical mean-field theory (DMFT), random phase approximation (RPA), and fluctuation exchange approximation (FLEX) methods to study its electronic structure and superconducting pairing symmetry. DFT band analysis indicates that the low-energy states are mainly dominated by three Ni-3$d_{x^2-y^2}$ orbitals with weak $d$-$p$ hybridization. Multi-orbital DMFT calculations yield an average $d_{x^2-y^2}$ filling of about 0.87, close to the optimal value for infinite-layer nickelates. By comparing the DMFT results of $d_{x^2-y^2}$ only and full Ni-3$d$ models, we find that their electronic properties are almost completely consistent, demonstrating the effectiveness of the simplified $d_{x^2-y^2}$ model. RPA and FLEX calculations based on this model consistently indicate a dominant $d_{x^2-y^2}$-wave pairing instability. These results establish La$_4$Ni$_3$O$_8$ as a system whose electronic structure closely resembles that of hole-doped cuprates, suggesting that the electronic pockets near the $\Gamma$- and $A$-points, which are characteristic of infinite-layer nickelates, may not be essential for superconductivity. The calculated superconducting transition temperature ($T_C$) is approximately 10\,K, in good agreement with experimental observations.
\end{abstract}

\maketitle

\emph{Introduction---}The discovery of superconductivity in nickelates has established a new family of unconventional superconductors beyond the cuprates \cite{li2019superconductivity,PhysRevLett.125.027001,PhysRevLett.125.147003,kitatani2020nickelate,held2022phase}. To date, superconductivity has been observed in three distinct classes of nickelates. The first class comprises infinite-layer nickelates $R$NiO$_2$ ($R$=La, Pr, Nd), where superconductivity emerges upon hole doping via divalent Ca or Sr substitution \cite{li2019superconductivity,osada2020phase,osada2021nickelate,zeng2022superconductivity,PhysRevLett.125.027001,PhysRevLett.125.147003}. Superconductivity has also been reported in nominally undoped NdNiO$_2$ \cite{PhysRevX.15.021048}.
Under high pressure, their superconducting transition temperature $T_C$ can be significantly enhanced, reaching above 70\,K at 90\,GPa \cite{lee2026high}. Spectroscopic experiments \cite{hepting2020electronic,goodge2021doping,ding2024cuprate} and many-body calculations \cite{PhysRevLett.126.127401,PhysRevB.101.020501,PhysRevB.101.041104,kitatani2020nickelate,held2022phase,PhysRevX.10.021061,PhysRevX.10.041002} have provided important insights into the orbital character and electronic correlations of infinite-layer nickelates. Superconducting electrodynamics measurements provide evidence consistent with nodal, cuprate-like $d$-wave pairing \cite{cheng2024evidence,PhysRevB.111.014519,fz8w-w61f}, while theoretical studies have also supported spin-fluctuation-mediated $d_{x^2-y^2}$-wave pairing mechanisms in infinite-layer nickelates \cite{PhysRevLett.125.077003,PhysRevB.106.035111,PhysRevB.109.235126}.
The second class is represented by the Ruddlesden-Popper (RP) phases La$_3$Ni$_2$O$_7$ \cite{sun2023signatures,hou2023emergence,zhang2024high} and La$_2$PrNi$_2$O$_{7}$ \cite{wang2024bulk}, as well as the trilayer compounds  La$_4$Ni$_3$O$_{10}$ \cite{zhu2024superconductivity,li2024signature} and Pr$_4$Ni$_3$O$_{10}$ \cite{pei2024pressure,zhang2025bulk}, in which superconductivity emerges under pressure without additional carrier (hole) doping. In these RP phases, the apical O brings the Ni-3$d_{z^2}$ and apical O-2$p_z$ orbitals to the Fermi level ($E_f$) \cite{PhysRevB.109.235123,PhysRevB.108.L140505,PhysRevLett.131.236002}, and most theoretical studies point to an $s^\pm$-wave pairing symmetry \cite{PhysRevLett.131.236002,PhysRevB.108.L140505,PhysRevB.109.L220506,PhysRevB.108.165141,PhysRevB.110.L180501,PhysRevLett.133.136001,zhang2024structural}. The third class consists of finite-layer structures, such as Nd(La)$_6$Ni$_5$O$_{12}$ \cite{pan2022superconductivity,shi2026bulk} and La$_5$Ni$_4$O$_{10}$ \cite{zhang2026ambient}. Subsequent theoretical studies revealed similarities between the electronic structures of quintuple-layer and infinite-layer nickelates and supported $d_{x^2-y^2}$-wave pairing in the broader $R_{n+1}$Ni$_n$O$_{2n+2}$ family \cite{hausoel2025superconducting,PhysRevMaterials.6.L091801,Wang2026Quintuple,she2026layer,qp4p-x6g3,yd8w-frs8}.

Very recently, signatures of ambient-pressure superconductivity were reported in Pr- and Y-substituted La$_4$Ni$_3$O$_8$ single crystals, with characteristic temperatures of approximately 7\,K and 15\,K, respectively \cite{ahn2026superconductivity}. This observation is particularly intriguing because its $T_C$ is comparable to that of infinite-layer nickelates \cite{li2019superconductivity,zeng2022superconductivity,kitatani2020nickelate,wei2023large}, despite the presence of trilayers of square planar NiO$_2$ sheets separated by fluorite-like spacer layers \cite{zhang2016stacked,zhang2017orbital,zhang2019spin,pardo2010quantum,pardo2012pressure}. Angle-resolved photoemission measurements \cite{Li2023ARPES} and resonant inelastic X-ray scattering (RIXS)   \cite{Lin2021Superexchange} have revealed cuprate-like Fermi-surface features and substantial electronic renormalization in Pr$_4$Ni$_3$O$_8$ and established a substantial antiferromagnetic superexchange in La$_4$Ni$_3$O$_8$, respectively.
Most importantly, neither simulations nor experiments in Ref.~\cite{Li2023ARPES} have revealed the $\Gamma$- and $A$-point electronic pockets in Pr$_4$Ni$_3$O$_8$. This raises two fundamental questions: how closely does (Pr,La)$_4$Ni$_3$O$_8$ resemble La(Nd)NiO$_2$, and are these electronic pockets essential for superconductivity? (Pr,La)$_4$Ni$_3$O$_8$ therefore provides an ideal platform for addressing these questions.

\begin{figure*}[!htbp]
\centering
\includegraphics[width=0.95\textwidth]{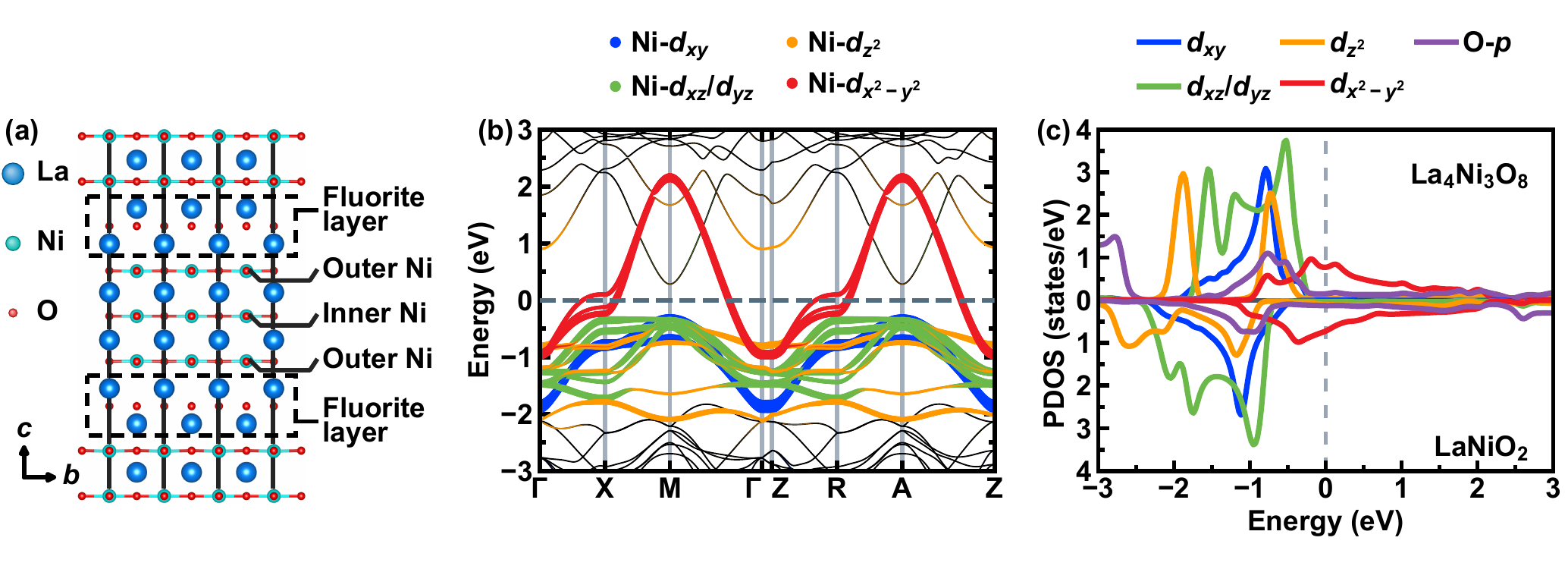}
\caption{Crystal and electronic structures of La$_4$Ni$_3$O$_8$, with LaNiO$_2$ as a reference.
(a) Crystal structure viewed along the $a$ axis, showing the inner and outer Ni sites. Dashed boxes identify the fluorite-like La$_2$O$_2$ spacer layers.
(b) Orbital-resolved non-spin-polarized DFT band structure of La$_4$Ni$_3$O$_8$, with colored weights indicating Ni-$3d$ character.
(c) Projected density of states (DOS) of La$_4$Ni$_3$O$_8$ (upper) and LaNiO$_2$ (lower).}
\label{fig:figure1}
\end{figure*}

Building on earlier studies of correlated electronic states and pairing in square-planar trilayer nickelates \cite{nica2020theoretical,karp2020comparative}, in this work, we perform a comprehensive theoretical study of La$_4$Ni$_3$O$_8$ by combining density-functional theory (DFT) \cite{PhysRev.136.B864,PhysRev.140.A1133}, dynamical mean-field theory (DMFT) \cite{PhysRevLett.62.324,RevModPhys.68.13}, random phase approximation (RPA) \cite{shao2026pairing,liu2013chiral}, and fluctuation-exchange (FLEX) \cite{kubo2007pairing,Bickers1989,Kitatani2022} calculations. The DFT band structure reveals that the low-energy states near the $E_f$ are dominated by the three Ni-3$d_{x^2-y^2}$ orbitals with only weak $d$-$p$ hybridization and without any pocket bands. By performing multi-orbital Ni-3$d$ DMFT calculations, we determine the $d_{x^2-y^2}$-orbital occupancies: $\sim$0.87 electrons per orbital, which is close to the optimal filling $n\sim$0.85 found in previous theoretical studies \cite{kitatani2020nickelate,held2022phase,PhysRevLett.130.166002}. A detailed analysis for the 3-band ($d_{x^2-y^2}$ only) and 15-band (Ni-3$d$) DMFT results  demonstrates that the minimal 3-band model captures the essential low-energy physics.

Using the tight-binding Hamiltonian derived from the $d_{x^2-y^2}$ 3-band Wannier projection, we perform RPA and FLEX calculations. The RPA results show that the $d_{x^2-y^2}$ pairing channel dominates for all interaction strengths considered. 
Logarithmic extrapolations of the FLEX pairing eigenvalues to $\lambda_{\mathrm{SC}}$=1 yield estimated pairing-instability temperatures of approximately 5.8\,K and 14.2\,K for $U$=1.0 and 1.1\,eV, respectively, in good agreement with the experiments. Therefore, our results establish that La$_4$Ni$_3$O$_8$ is a hole-doped Hubbard superconductor with $d_{x^2-y^2}$-wave symmetry pairing, and that its minimal electronic structure consists of three $d_{x^2-y^2}$ bands, closely analogous to infinite-layer nickelate superconductors. Most importantly, the electronic pockets around the $\Gamma$- and $A$-points in infinite-layer nickelates are not essential for superconductivity.

\emph{Methods---}The DFT-level electronic structures were calculated using the
\verb|Vasp| \cite{Kresse1996PRB,Kresse1996CMS} and \verb|Wien2K| \cite{Blaha2001WIEN2k,schwarz2002electronic,Blaha2020} codes. The exchange-correlation functional was treated using the Perdew-Burke-Ernzerhof version of the generalized gradient approximation (GGA-PBE) \cite{Perdew1996PRL}. The experimental lattice parameters of La$_4$Ni$_3$O$_8$, $a$=$b$=3.97\,\AA~and $c$=26.09\,\AA~were adopted for calculations \cite{zhang2016stacked}. Brillouin-zone samplings were performed on a dense
16$\times$16$\times$4 $k$-mesh. The corresponding DFT bands were projected onto maximally localized Wannier functions
\cite{RevModPhys.84.1419} to construct low-energy effective Hamiltonians using the \verb|Wannier90| code \cite{mostofi2008wannier90} and the \verb|Wien2Wannier| interface \cite{kunevs2010wien2wannier}. Two models were constructed for La$_4$Ni$_3$O$_8$: (1) a 15-band model including all Ni-$3d$ orbitals of all the three Ni atoms (in a unit cell); and (2) a 3-band model containing only the Ni-$d_{x^2-y^2}$ orbital of all Ni atoms.

The resulting tight-binding Hamiltonians were supplemented with local electronic interactions, and the fully localized limit was employed for the double-counting correction \cite{liechtenstein1995density,anisimov1993density}. The interaction parameters for 15-band DMFT calculations (intra-orbital $U$=4.4\,eV, Hund's exchange $J$=0.65\,eV and inter-orbital $U'$=3.1\,eV) were adopted from Refs.~\cite{PhysRevLett.130.166002,PhysRevLett.124.166402}. For the 3-band model, a local Hubbard interaction of $U$=3.2\,eV was used for DMFT calculations. The RPA calculations were performed at $U$=0.6, 0.8,
and 1.0\,eV, whereas the FLEX calculations covered $U$=0.6-1.4\,eV. Larger interaction strengths were excluded due to convergence difficulties. The resulting many-body Hamiltonians were solved in the paramagnetic state at room temperature $T$=300\,K within the dynamical mean-field theory (DMFT) framework \cite{RevModPhys.68.13}, using the continuous-time quantum Monte Carlo impurity solver in the hybridization expansion \cite{gull2011continuous}, as implemented in the \verb|W2dynamics| code \cite{parragh2012conserved,wallerberger2019w2dynamics}. Real-frequency spectral functions were obtained via analytic continuation using the maximum entropy method (MaxEnt) \cite{gubernatis1991quantum}, as implemented in the \verb|ana_cont| code~\cite{kaufmann2023ana_cont}.

To investigate magnetic fluctuations and superconducting pairing, we consider a 3-band Hubbard model constructed from the
Ni-$d_{x^2-y^2}$ Wannier orbitals,

\begin{equation}
\begin{aligned}
H_{\mathrm{3b}}=&\sum_{\mathbf{k},\sigma}\sum_{\ell,\ell'=1}^{3}c_{\mathbf{k}\ell\sigma}^{\dagger}\left[h_{\ell\ell'}(\mathbf{k})-\mu\delta_{\ell\ell'}\right]c_{\mathbf{k}\ell'\sigma}\\&+U\sum_{i,\ell}n_{i\ell\uparrow}n_{i\ell\downarrow}.
\end{aligned}
\end{equation}

Here, $\ell$ labels the three Ni sites within a unit cell, each represented by one $d_{x^2-y^2}$ Wannier orbital, and $i$ labels the unit cell. The Wannier Hamiltonian $h_{\ell\ell'}(\mathbf{k})$ contains the onsite energies and intra- and interlayer hoppings. The interaction $U$ acts locally on each Ni orbital, while the chemical potential $\mu$ is adjusted to maintain the prescribed total filling. We treat this model within the random-phase approximation (RPA) \cite{graser2009near,PhysRevLett.133.136001,shao2026pairing,liu2013chiral} and the self-consistent fluctuation-exchange approximation (FLEX) \cite{Bickers1989,qp4p-x6g3,Kitatani2022}. Guided by the Ni-3$d_{x^2-y^2}$ orbital occupations obtained from the 15-band DMFT calculations, the total electron filling in the RPA calculations was fixed at $N$=2.60 per unit cell. Brillouin-zone integrations were performed on a 144$^3$ $k$-mesh, while a 48$^3$ mesh was employed for the susceptibility and pairing calculations.
The static spin susceptibility and Fermi-surface pairing strengths $\lambda_{\mathrm{FS}}$ were evaluated for different pairing symmetry channels. For the FLEX calculations, the spin-summed average filling was fixed at $n$=0.87 electrons per orbital, using a 128$\times$128$\times$1 momentum mesh. The electronic self-energy and spin susceptibility were calculated self-consistently, and the linearized Eliashberg equation was solved over a range of interaction strengths and temperatures to determine the leading pairing eigenvalues and eigenfunctions in the $d_{x^2-y^2}$-wave channel. All electron fillings include both spin components.

\emph{DFT band structure---}The crystal structure of La$_4$Ni$_3$O$_8$ is shown in Fig.~\ref{fig:figure1}(a). Each trilayer consists of one inner and two outer NiO$_2$ planes. In a unit cell, there are two inequivalent Ni sites: the outer Ni is adjacent to the interface La$_2$O$_2$ fluorite-layer while the inner Ni is located inside. Their inequivalent local environments can lead to different orbital on-site potentials, occupancies and correlations. The orbital-resolved non-spin-polarized DFT band structure [Fig.~\ref{fig:figure1}(b)] shows that the bands crossing the $E_f$ are mainly of Ni-3$d_{x^2-y^2}$ character, while the remaining Ni-3$d$ and O-2$p$ orbitals lie mostly below the $E_f$. Compared with infinite-layer nickelates \cite{kitatani2020nickelate}, the most notable difference is that at the $\Gamma$-, $M$- and $A$-points, the pocket-like band structures shift above the $E_f$, which makes the electron occupancy of the La-5$d$ and 4$f$ orbitals almost vanish. Meanwhile, the nearly identical band dispersions along $\Gamma$–$X$–$M$–$\Gamma$ and $Z$–$R$–$A$–$Z$ indicate a quasi-two-dimensional low-energy electronic structure. However, the interlayer hybridization of the $d_{x^2-y^2}$ orbitals gives rise to low-energy bands of the $d_{x^2-y^2}$ orbitals from the three NiO$_2$ planes (especially at the $X$- and $R$-points), which highlights the importance of retaining interlayer coupling in a simplified low-energy description.

To further understand the electronic structure of La$_4$Ni$_3$O$_8$, especially its difference from the infinite-layer nickelate LaNiO$_2$ \cite{PhysRevX.10.011024,Jiang2019RNiO2}, we calculated the orbital-resolved density of states [Fig.~\ref{fig:figure1}(c)]. 
For La$_4$Ni$_3$O$_8$, we choose the inner Ni as representative. It can be seen that the two materials have very similar electronic structures: the electronic structure near the $E_f$ is almost entirely contributed by Ni-3$d_{x^2-y^2}$. A notable difference is that in La$_4$Ni$_3$O$_8$, other electronic states, including other Ni-3$d$ and O-2$p$ orbitals, lie closer to the $E_f$. This is because La$_4$Ni$_3$O$_8$ has a larger in-plane lattice constant than LaNiO$_2$ ($\sim$3.95\,\AA)~\cite{PhysRevMaterials.7.014804}, leading to reduced bandwidth. Compared with cuprates such as CaCuO$_2$ \cite{PhysRevX.10.011024}, the in-plane O-2$p_x$/$p_y$ orbitals' contribution to the electronic states near the $E_f$ is negligible in both La$_4$Ni$_3$O$_8$ and LaNiO$_2$. Therefore, the O-2$p$ orbitals can be excluded from the minimal model that is a $d_{x^2-y^2}$ Hubbard model rather than an Emery model \cite{PhysRevLett.58.2794} including in-plane O-2$p_x$/$p_y$. 
However, this predominantly Ni-$d_{x^2-y^2}$ character does not imply that the Ni-O hybridization can be completely  excluded in spectroscopy such as RIXS \cite{de2024resonant}: O $K$-edge RIXS measurements combined with cluster calculations have
identified appreciable ligand-hole character in La$_4$Ni$_3$O$_8$ \cite{shen2022role}. Ni $L_2$-edge RIXS further revealed that charge order in La$_4$Ni$_3$O$_8$ involves predominantly Ni-$d_{x^2-y^2}$ and O-$2p_\sigma$ states, with a smaller contribution from Ni-$d_{3z^2-r^2}$ \cite{Shen2023ChargeOrder}.

\begin{figure*}[!htbp]
\centering
\includegraphics[width=0.9\textwidth]{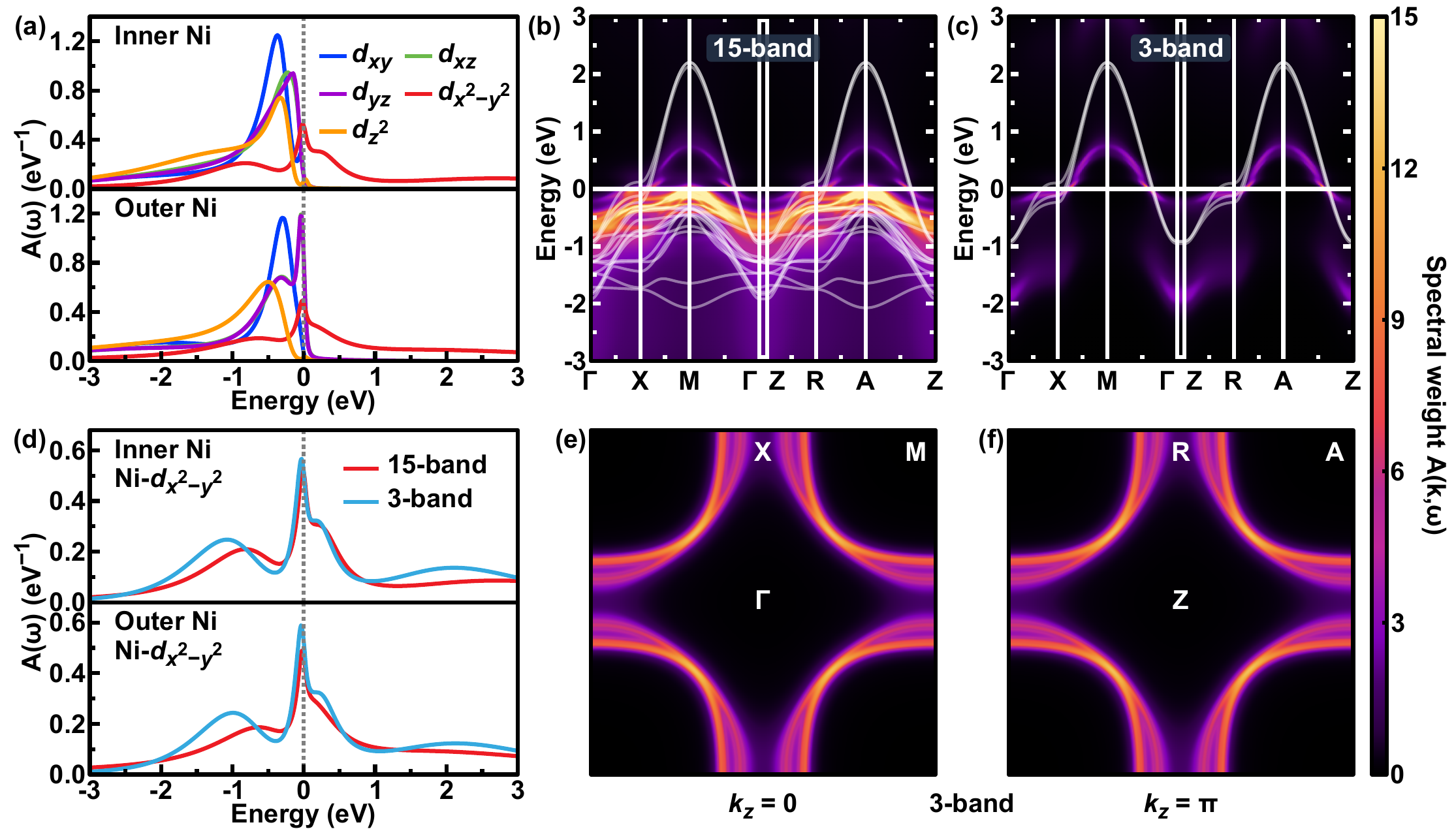}
\caption{DMFT spectral functions of the 15-band and 3-band models.
(a) Orbital-resolved local spectral functions for the inner-layer
(top) and outer-layer (bottom) Ni sites in the 15-band model.
(b,c) Total momentum-resolved spectral functions of the 15-band
(b) and 3-band (c) models. White curves indicate the corresponding
noninteracting reference bands.
(d) Comparison of the local Ni-$d_{x^2-y^2}$ spectral functions
between the 15-band (red) and 3-band (blue) models
for the inner-layer (top) and outer-layer (bottom) Ni sites.
(e,f) Zero-energy spectral weight of the 3-band model
in the $k_z$=0 (e) and $k_z$=$\pi$ (f) planes.}
\label{fig:figure2}
\end{figure*}

\emph{DMFT Corrections---}Strong electron correlation effects can significantly renormalize band structure, especially near $E_f$. We therefore employ the DMFT method to calculate the spectral functions of La$_4$Ni$_3$O$_8$. In  Fig.~\ref{fig:figure2}(a), the 15-band DMFT calculation using all Ni-3$d$ orbitals shows that in La$_4$Ni$_3$O$_8$ the Fermi surface contribution is mainly from $d_{x^2-y^2}$. However, compared with the DFT bands [Fig.~\ref{fig:figure2}(b)], the other Ni-3$d$ bands in DMFT shift significantly upward and approach the $E_f$. This is because strong correlation effects induce a slight transfer of electrons from other Ni-3$d$ orbitals to Ni-3$d_{x^2-y^2}$.

Specifically, see Table~\ref{tab:occupations}, in the 15-band DFT calculation, the Ni-3$d_{x^2-y^2}$ occupations of inner and outer Ni are 0.72 and 0.67, respectively, which are significantly smaller than the computed optimal superconducting value 0.85 and the minimum occupation 0.79 in infinite-layer nickelate superconductors \cite{kitatani2020nickelate}. However, in the 15-band DMFT calculation, due to strong correlation effects, about 0.1-0.2 electrons transfer from other Ni-3$d$ orbitals to $d_{x^2-y^2}$, making the $d_{x^2-y^2}$ occupations 0.88 (outer Ni) and 0.86 (inner Ni). These values are very close to the optimal value of 0.85 in infinite-layer nickelates \cite{kitatani2020nickelate,PhysRevLett.130.166002}. Therefore, the electron transfer caused by strong correlations enables La$_4$Ni$_3$O$_8$ to exhibit superconductivity even without extra hole doping. From the 15-band DMFT calculation, we find that the $d_{x^2-y^2}$ orbitals of the three Ni sites (two outer Ni and one inner Ni) are occupied by about 2.6 electrons in total. Figure~\ref{fig:figure2}(c) shows the momentum-resolved spectral function $A$($\mathbf{k}, \omega$) of the 3-band model, while Fig.~\ref{fig:figure2}(d) compares its local $d_{x^2-y^2}$ spectra with those of the 15-band model. Both models retain a low-energy $d_{x^2-y^2}$ peak and broader occupied-state spectral features, although their peak intensities, spectral-weight distributions, and inner--outer layer differentiation remain different. These similarities support the 3-band model as an effective description of the low-energy $d_{x^2-y^2}$ sector.
Moreover, we further estimate the orbital-resolved mass enhancements from 
\[
Z_{\alpha}^{-1}
=
1-
\left.
\frac{\partial\,\mathrm{Im}\Sigma_{\alpha}(i\omega)}
{\partial\omega}
\right|_{\omega\rightarrow0^{+}},
\]
where $\alpha$ labels the orbital. Using the above equation, we obtain orbital renormalization estimates $Z_{\alpha}^{-1}$ of 3.77 and 3.88 for the inner and outer Ni-3$d_{x^2-y^2}$ orbitals in the 15-band model, respectively, compared with 3.75 and 3.43 in the 3-band model. The effective masses obtained from the two different models are nearly identical, demonstrating that the $d_{x^2-y^2}$ orbital can be treated as an independent orbital and that no Hund's-metal-like multi-orbital effect \cite{PhysRevLett.126.127401,PhysRevB.101.041104,wang2020hund} emerges between it and the other Ni-3$d$ orbitals.
In Figs.~\ref{fig:figure2}(e,f), we plot the Fermi surface from the 3-band DMFT calculations (at $k_z$=0 and $\pi$). It can be seen that the Fermi surface exhibits clear hole-doped two-dimensional characteristics. This is because in the $z$-direction, the hopping between the two $d_{x^2-y^2}$ orbitals separated by the interface fluorite La$_2$O$_2$ layer is only 0.67\,meV (0.35\,meV) in the Wannier projection for the 3-band (15-band) model.
Hence, from the DMFT calculations of the 3-band and 15-band models, the minimal model of La$_4$Ni$_3$O$_8$ can be represented by three $d_{x^2-y^2}$ orbitals with quasi-two-dimensional characteristics even under strong correlation effects. In fact, it is precisely the electron transfer behavior caused by strong correlation effects that drives the $d_{x^2-y^2}$ occupation into the optimal (superconducting) region, leading to the superconductivity in La$_4$Ni$_3$O$_8$, similar to that in Nd$_6$Ni$_5$O$_{12}$ \cite{PhysRevMaterials.6.L091801}. 
Finally, the occupation of the three $d_{x^2-y^2}$ orbitals was extracted from the 15-band DMFT calculations, yielding a total filling of $N$=2.60, corresponding to an average $n$=0.87 electrons per $d_{x^2-y^2}$. This filling was then used in the subsequent RPA and FLEX calculations.

\begin{table}[t]
\caption{Spin-summed orbital occupations for the inner and outer Ni sites in the 15-band and 3-band models of La$_4$Ni$_3$O$_8$, obtained from DFT and DMFT calculations. The 3-band model retains only the Ni-$d_{x^2-y^2}$ orbital.}
\label{tab:occupations}
\centering
\small
\setlength{\tabcolsep}{3.5pt}
\renewcommand{\arraystretch}{1.2}
\begin{tabular}{c|c|c|c|c|c|c}
\hline
Model & Site & Method & $d_{xy}$ & $d_{xz/yz}$ & $d_{x^2-y^2}$ & $d_{z^2}$ \\
\hline

\multirow{4}{*}{15-band}
& \multirow{2}{*}{Inner-Ni}
& DFT  & 2.00 & 2.00 & 0.72 & 1.98 \\
\cline{3-3}
&
& DMFT & 1.98 & 1.97 & 0.86 & 1.98 \\
\cline{2-3}

& \multirow{2}{*}{Outer-Ni}
& DFT  & 2.00 & 2.00 & 0.67 & 1.99 \\
\cline{3-3}
&
& DMFT & 1.98 & 1.89 & 0.88 & 1.99 \\
\hline

\multirow{4}{*}{3-band}
& \multirow{2}{*}{Inner-Ni}
& DFT  & -- & -- & 0.90 & -- \\
\cline{3-3}
&
& DMFT & -- & -- & 0.89 & -- \\
\cline{2-3}

& \multirow{2}{*}{Outer-Ni}
& DFT  & -- & -- & 0.85 & -- \\
\cline{3-3}
&
& DMFT & -- & -- & 0.86 & -- \\
\hline

\end{tabular}
\end{table}

\begin{figure}[!htbp]
\centering
\includegraphics[width=0.95\columnwidth]{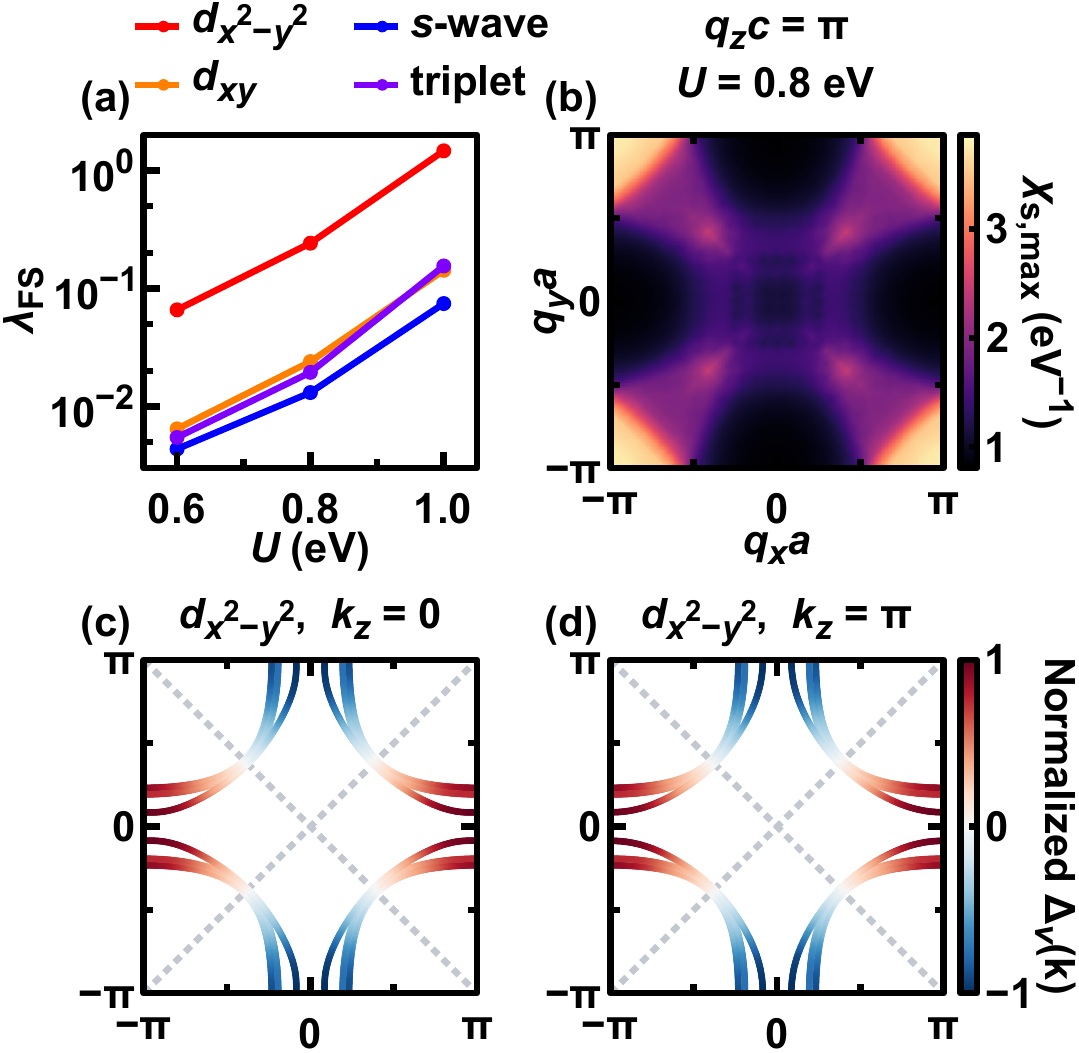}
\caption{Pairing channels and spin response of the 3-band $d_{x^2-y^2}$ model obtained within RPA. The calculations are performed at a fixed total filling of $N$=2.60 and a temperature of $T$=69.6\,K (corresponding to $k_{\mathrm{B}}T$=0.006\,eV).
(a) Static Fermi-surface pairing strength, $\lambda_{\mathrm{FS}}$, as a function of $U$ for different pairing channels.
(b) Largest eigenvalue of the static spin-susceptibility matrix in the $q_zc$=$\pi$ plane at $U$=0.8\,eV.
(c,d) Normalized pairing eigenfunction on the Fermi-surface cross-sections at $k_z$=0 (c) and $k_z$=$\pi$ (d) for the same interaction strength, showing the sign changes and nodes characteristic of $d_{x^2-y^2}$ pairing.}
\label{fig:figure3}
\end{figure}

\emph{Competing pairing channels from RPA calculations---}We first examine the competition among different pairing symmetries  by employing the RPA method. Figure~\ref{fig:figure3}(a) shows the static Fermi-surface pairing eigenvalues $\lambda_{\mathrm{FS}}$ as a function of the interaction strength $U$ at a fixed total filling of $N$=2.60 for the 3-band $d_{x^2-y^2}$ model, and a temperature of $T$=69.6\,K (corresponding to $k_{\mathrm{B}}T$=0.006\,eV). At all three interaction values studied, the $d_{x^2-y^2}$ pairing channel has the largest pairing eigenvalue. Its (static) Fermi-surface pairing strength increases from approximately 0.066 at $U$=0.6\,eV to 1.466 at $U$=1.0\,eV, while among the other competing channels, the largest strength  merely  reaches $\sim$0.155, indicating that the $d_{x^2-y^2}$ pairing is the leading channel among those examined. This pairing symmetry is also favored in earlier $t$--$J$-model calculations for doped square-planar trilayer nickelates \cite{nica2020theoretical}.
With increasing $U$, spin fluctuations in La$_4$Ni$_3$O$_8$ are enhanced and the pairing strength in the $d_{x^2-y^2}$ channel increases rapidly. At $U$=1\,eV, the Stoner factor is approximately 0.944, which starts to approach the RPA magnetic instability; therefore, in the RPA calculations we only considered $U$ up to 1\,eV. The $\lambda_{FS}$ shown in Fig.~\ref{fig:figure3} is used to compare the pairing strengths of different channels and cannot be used to directly determine the superconducting transition temperature. In the following, we employ the FLEX method and solve the Eliashberg equation to obtain $T_C$ .

The momentum-dependence of the spin response provides a physical interpretation of the leading pairing symmetry. At $U$=0.8\,eV, the largest eigenvalue of the static spin-susceptibility matrix is enhanced near the in-plane wave vector ($\pi$/a,$\pi$/a) in the $q_zc$=$\pi$ plane [see Fig.~\ref{fig:figure3}(b)]. A repulsive spin-fluctuation-mediated pairing interaction favors opposite gap signs on Fermi-surface regions connected by the dominant scattering wave vectors \cite{graser2009near}. The $d_{x^2-y^2}$ form factor, $\cos(k_xa)$-$\cos(k_ya)$, satisfies this in-plane sign-change condition. Consistently, the calculated pairing eigenfunctions exhibit opposite signs along the two orthogonal in-plane directions and nodes near $k_x$=$\pm k_y$ in both the $k_z$=0 and $k_z$=$\pi$ sections [Figs.~\ref{fig:figure3}(c) and (d)]. These results connect the leading $d_{x^2-y^2}$ pairing channel with the enhanced antiferromagnetic spin response.

\begin{figure}[!htbp]
\centering
\includegraphics[width=1 \columnwidth]{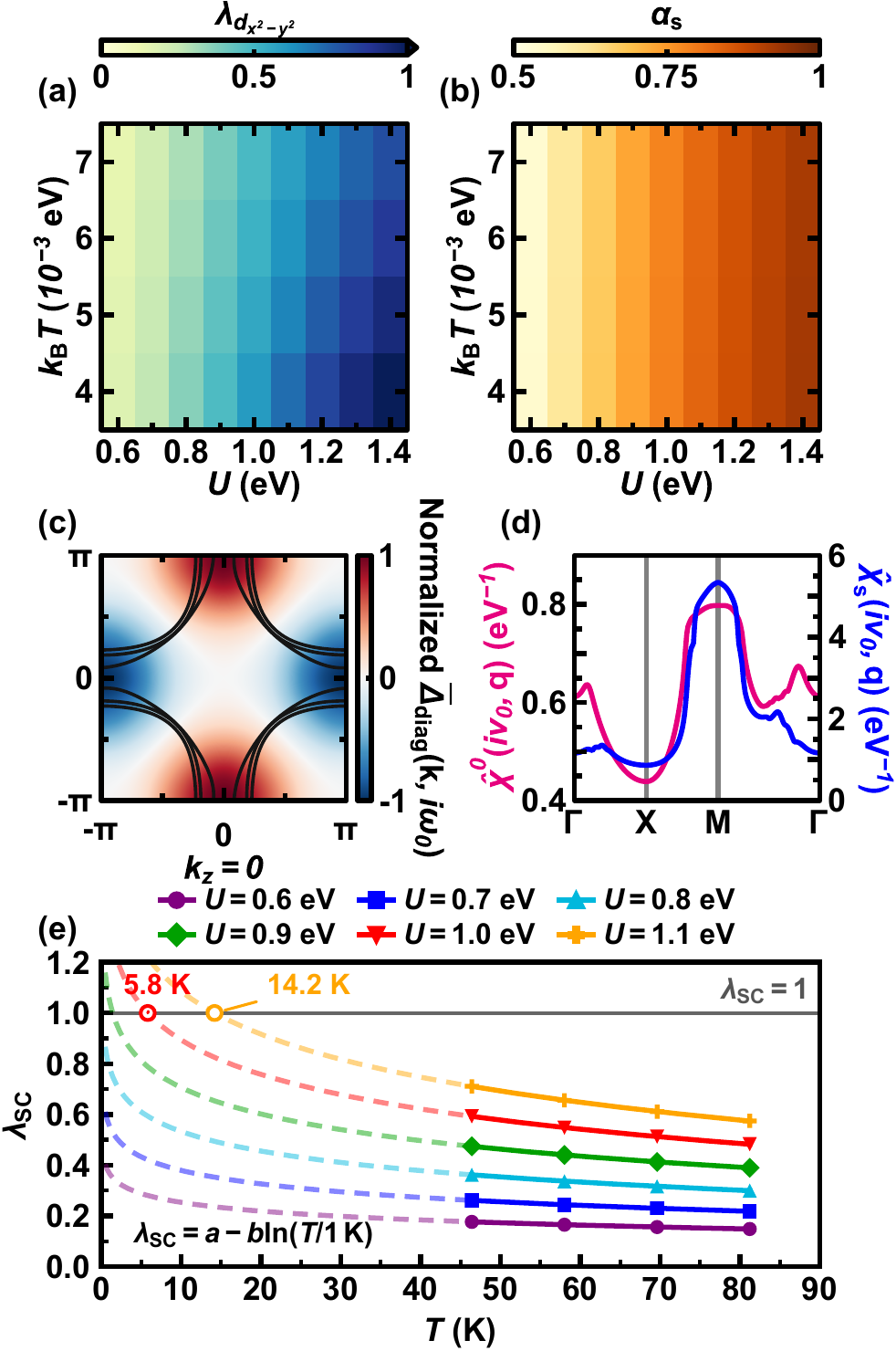}
\caption{FLEX results for the $d_{x^2-y^2}$ 3-band model.
(a) $d_{x^2-y^2}$ pairing eigenvalue $\lambda_{\mathrm{SC}}$ and (b) Stoner factor $\alpha_S$ as functions of interaction $U$ and $k_BT$.
(c) Momentum-dependence of the normalized $d_{x^2-y^2}$ pairing eigenfunction. Black lines indicate the Fermi-surface contours obtained from FLEX.
(d) Static susceptibilities $\chi_0$ (left axis) and $\chi_s$ (right axis) along the $\Gamma$--$X$--$M$--$\Gamma$ path.
(e) Superconducting eigenvalue $\lambda_{\mathrm{SC}}$ vs.~temperature $T$. The superconducting transition temperature $T_C$ corresponds to $\lambda_{\mathrm{SC}}\to$1.}
\label{fig:figure4}
\end{figure}

\emph{$d_{x^2-y^2}$-wave pairing and spin fluctuations within FLEX---}By self-consistently updating the electronic self-energy, Green's function, and fluctuation-mediated interactions, 
the FLEX approximation provides a self-consistent treatment of self-energy and frequency-dependent pairing interaction, both constructed consistently from the same underlying spin and charge fluctuation propagators \cite{Bickers1989,Monthoux1994,Takimoto2004,qp4p-x6g3,PhysRevLett.125.077003,PhysRevB.103.205148}. Hence, to examine the effect of correlations and dynamical fluctuations on the superconducting pairing tendency, we next perform (self-consistent) FLEX calculations for the $d_{x^2-y^2}$ 3-band model and solve the linearized Eliashberg equation in the $d_{x^2-y^2}$-wave channel at a spin-summed filling of $n$=0.87 electrons per $d_{x^2-y^2}$ orbital. The pairing eigenvalue in this channel, $\lambda_{\mathrm{SC}}$, increases with increasing $U$ and decreasing temperature [Fig.~\ref{fig:figure4}(a)]. The Stoner factor $\alpha_S$ also increases with $U$, but exhibits a much weaker temperature dependence [Fig.~\ref{fig:figure4}(b)]. For example, at $U$=1.4\,eV, a cooling process from 81.2\,K (corresponding to $k_{\mathrm{B}}T$=0.007\,eV) to 46.4\,K (0.004\,eV) increases $\lambda_{\mathrm{SC}}$ from approximately 0.80 to 1.02, achieving a superconducting state, whereas $\alpha_S$ changes only from 0.918 to 0.924. Thus, pairing is substantially enhanced upon cooling, while the Stoner factor remains nearly unchanged.

The decoupling between the pairing eigenvalue $\lambda$ (rising above 1) and the nearly constant Stoner factor $\alpha_S$ should be interpreted with the intrinsic shortcoming of FLEX \cite{Kontani2021}. FLEX is a spin-fluctuation-based RPA-type theory that neglects vertex corrections \cite{Onari2012}; these corrections are often minor for the spin channel but can be significant for the charge and orbital channels, so the charge susceptibility may be underestimated. Consequently,  $\alpha_S$ primarily measures the proximity to a static magnetic instability in the spin channel and does not capture all fluctuations relevant to pairing. If part of the pairing glue arises from charge, orbital-polarization, or orbital-selective fluctuations, the effective pairing interaction can grow with decreasing temperature $T$ even when $\alpha_S$ remains nearly unchanged, pushing $\lambda$ above 1. Thus, a constant $\alpha_S$ does not rule out a magnetic contribution, but indicates that $\alpha_S$ is not the only/sufficient indicator of pairing enhancement, at least for the La$_4$Ni$_3$O$_8$. To test this, one should also compute the charge susceptibility $\chi_c$($q$,$\omega$), orbital-resolved susceptibilities, and vertex corrections, compare with DFT+DMFT or other advanced methods such as multi-band D$\Gamma$A \cite{Galler2017,Rohringer2018}, and analyze the orbital and momentum structure of the pairing gap $\Delta$($k$) for further investigations.

The momentum structure of the pairing eigenfunction and spin response is shown in Figs.~\ref{fig:figure4}(c,d) for $U$=1.2\,eV and $T$=69.6\,K. Figure~\ref{fig:figure4}(c) shows the momentum dependence of the pairing function, with the Fermi-surface contours superimposed. The pairing function changes sign under $k_x\leftrightarrow k_y$ and has nodes along the Brillouin-zone diagonals, displaying the $d_{x^2-y^2}$ structure also found in RPA. Figure~\ref{fig:figure4}(d) shows the momentum dependence of the static irreducible susceptibility $\chi_0(\mathbf{q},0)$ and spin susceptibility $\chi_s(\mathbf{q},0)$ along $\Gamma$--$X$--$M$--$\Gamma$. At each momentum $\mathbf{q}$, we plot the largest eigenvalue of the corresponding susceptibility matrix. The spin response exhibits a pronounced peak near $M$=($\pi/a$,$\pi/a$), indicating enhanced antiferromagnetic spin fluctuations at this wave vector, consistent with the sign-changing $d_{x^2-y^2}$ pairing structure in Fig.~\ref{fig:figure4}(c).

Finally, to estimate the superconducting critical temperature $T_C$, Fig.~\ref{fig:figure4}(e) shows $\lambda_{\mathrm{SC}}(T)$ for $U$=0.6\,eV-1.1\,eV. The eigenvalue $\lambda_{\mathrm{SC}}$ increases with both decreasing $T$ and increasing $U$. We fit the temperature dependence of $\lambda_{\mathrm{SC}}$ using $\lambda_{\mathrm{SC}}(T)$=$a$-$b\ln(T/\mathrm{K})$, with $a$ and $b$ fitted separately for each $U$, and extrapolate to $\lambda_{\mathrm{SC}}$=1. For $U$=1.0\,eV and 1.1\,eV, this yields estimated pairing-instability temperatures of approximately 5.8\,K and 14.2\,K, respectively. These estimates are on the same temperature scale as the superconducting onsets reported near 7\,K and 15\,K in Pr- and Y-substituted La$_4$Ni$_3$O$_8$ \cite{ahn2026superconductivity}.

\emph{Conclusion and outlook---}We investigated the correlated electronic structure and superconducting pairing tendencies of La$_4$Ni$_3$O$_8$ using DFT, DMFT, RPA, and FLEX. Unlike infinite-layer nickelates, La$_4$Ni$_3$O$_8$ exhibits no electronic pockets around $\Gamma$- and $A$-points. Its low-energy states near the $E_f$ are dominated by the three Ni-3$d_{x^2-y^2}$ orbitals, and the weak $d$-$p$ hybridization justifies a minimal three-band model with one Ni-3$d_{x^2-y^2}$ Wannier orbital per Ni. The 15-band and 3-band DMFT calculations give nearly identical effective masses, Fermi surfaces, and spectral functions. Correlations redistribute the $d_{x^2-y^2}$ occupancy from 0.67-0.72 in DFT to about 0.87 electrons per orbital in DMFT, close to the optimal filling of infinite-layer nickelates. At this filling, the RPA calculation identifies the $d_{x^2-y^2}$-wave channel as the leading pairing channel, and FLEX shows that the pairing in this channel strengthens with increasing (decreasing) interaction $U$  (temperature $T$). The FLEX pairing eigenfunction changes sign on the Fermi surface and is consistent with the enhanced antiferromagnetic spin response around ($\pi$,$\pi$). Extrapolating the FLEX eigenvalues gives pairing instability temperatures of about 5.8\,K and 14.2\,K for $U$=1.0\,eV and 1.1\,eV, comparable to the experimental 7\,K-15\,K. Since FLEX neglects vertex corrections and tends to overestimate the pairing strength and $T_C$, the RPA/FLEX results are mainly used to identify the intrinsic pairing channel and estimate the $T_C$ scale; accurate quantitative predictions require other methods such as D$\Gamma$A \cite{Rohringer2018,PhysRevLett.130.166002,wu2026single} or DCA \cite{RevModPhys.77.1027,PhysRevB.88.115101,PhysRevB.61.12739}. Finally, our results indicate that the $\Gamma$- and $A$-pockets in infinite-layer nickelates are not essential for nickelate superconductivity. La$_4$Ni$_3$O$_8$ is an unconventional superconductor whose pairing mechanism is driven by antiferromagnetic spin fluctuations, exhibiting characteristics associated with hole-doped cuprates.

Given the similarities in the electronic structures and leading superconducting pairing tendencies between La$_4$Ni$_3$O$_8$ and LaNiO$_2$, the relatively large in-plane lattice constant of La$_4$Ni$_3$O$_8$ (3.97\,\AA) \cite{poltavets2007crystal,PhysRevLett.104.206403}, compared with those of LaNiO$_2$ (3.95\,\AA) \cite{PhysRevMaterials.7.014804} and LaNiO$_2$ films grown on SrTiO$_3$ substrates (3.905\,\AA), suggests an opportunity for strain engineering. 
Recent experiments have reported enhanced $T_C$ in infinite-layer nickelates under external pressure \cite{wang2022pressure,lee2026high,cai2026enhanced,4h3v-tjwz}. Theoretical studies have suggested that this enhancement is associated with the (pressure-induced) enhanced $d_{x^2-y^2}$ bandwidth and the reduced effective correlation strength \cite{di2024unconventional}, characterized by the reduced $U/t$ ratio \cite{PhysRevLett.130.166002}.
We hence propose that epitaxial growth of La$_4$Ni$_3$O$_8$ on substrates with smaller in-plane lattice constants, such as LSAT (3.868\,\AA) \cite{tidrow1997new} or LaAlO$_3$ (3.78\,\AA) \cite{rizwan2019review}, could modify the Ni-3$d_{x^2-y^2}$ bandwidth and magnetic fluctuations, potentially enhancing the superconducting pairing instability while retaining sufficiently strong antiferromagnetic spin fluctuations. This possibility warrants further investigation, particularly considering the potential competition between superconductivity and the stripe-ordered insulating state in La$_4$Ni$_3$O$_8$ \cite{zhang2019spin,zhang2016stacked}.

\emph{Acknowledgments---}R.~Y.~and Z.~X.~S.~acknowledge support from the National Key R\&D Program of China (Grant No.~2024YFA1408400) and the National Natural Science Foundation of China (Grant Nos.~12374135 and 12474152). L.~S.~acknowledges support from the National Natural Science Foundation of China (Grant Nos.~12422407 and 12674309) and the Key Research and Development Program of Shaanxi (2024QY2-GJHX-42). 
Calculations have been done on the National Supercomputing Center in Northwest University (Xi'an). 

\emph{Data availability---}The data that support the findings of this article are not publicly available. The data are available from the authors upon reasonable request.

\end{document}